\documentclass{article}
\usepackage[left=3cm,right=3cm,top=2cm,bottom=2cm]{geometry} 
\usepackage{amsmath} 
\usepackage{caption}
\usepackage{listings}
\usepackage{caption}
\usepackage{cite}
\usepackage{graphicx}
\usepackage{subfig}
\usepackage{epstopdf}
\usepackage{booktabs}
\usepackage{multirow}
\usepackage{mwe}
\usepackage{color, colortbl}
\usepackage[section]{placeins}
\begin{document}
\title{Influence of Polarization Transformation in Phase Conjugation of Polarization Multiplexed Data using Bragg-Scattering FWM in nonlinear SOAs}
\author{\textbf{Aneesh Sobhanan and Deepa Venkitesh$^*$}\\ Department of Electrical Engineering\\ Indian Institute of Technology Madras, India-600036\\ $^*$\textit{deepa@ee.iitm.ac.in}}
\date{}
\maketitle
\section*{Abstract}
\footnotesize We theoretically investigate the suitability of the selection of pump frequency ($\omega_p$)  with respect to the signal frequency ($\omega_s$) for the best polarization insensitive phase conjugate generation through Bragg-scattering based four-wave mixing in nonlinear SOAs. We study the positive detuning ($\omega_s>\omega_p$) and negative detuning ($\omega_s<\omega_p)$ scenario for the inherent polarization transformation of conjugate generation at both the ports of SOA. Conjugate generation is desired to be through a unitary transformation for its utility in further digital signal processing, and we prove that only positive detuning conditions result in unitary transformation.
\section{Introduction}
The mechanism of polarization insensitive optical phase conjugate (OPC) generation using Bragg-Scattering four-wave mixing (BS-FWM) in both the ports of the SOA in partially degenerate configuration is explained in our previous work~\cite{Sobhanan}. We discuss the phase matching conditions that lead to efficient conjugate generation, along with the corresponding allowed and forbidden mixing processes~\cite{Sobhanan}. In the same paper, we also provide experimental evidence of polarization insensitive phase conjugation with CW signals.  In this work, we present the specific polarization transformation properties of conjugate generation process in order to optimally allocate the relative frequency of the pump with respect to the signal. Unitary transformation is highly desirable for its utility in the most commonly used digital signal processing algorithms~\cite{Ip}.
\section{Analytical formulation}
\begin{figure}
	\centering
	\subfloat[]
	{
		\includegraphics[scale=.5]{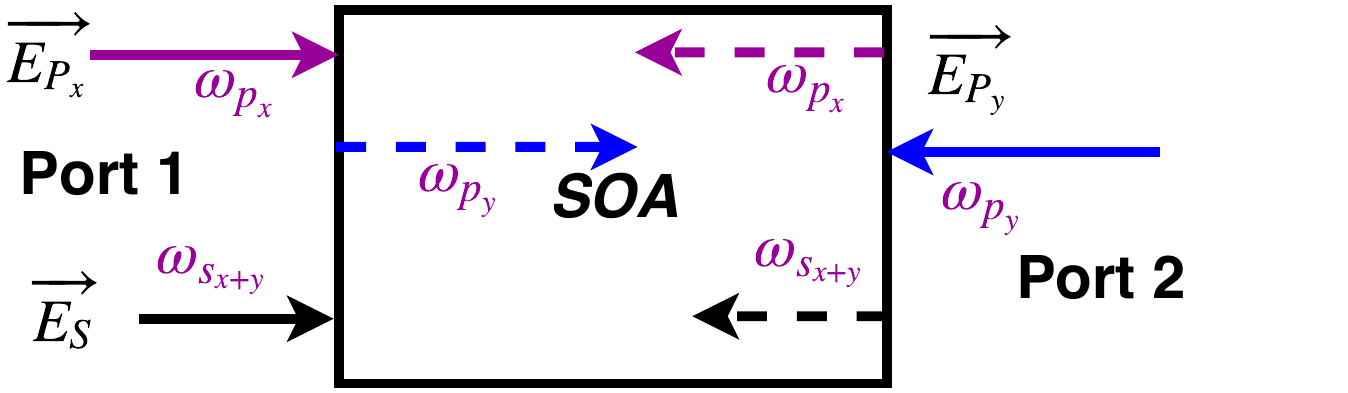}
		\label{scheme1}
	}
	\subfloat[]
	{
		\includegraphics[scale=.5]{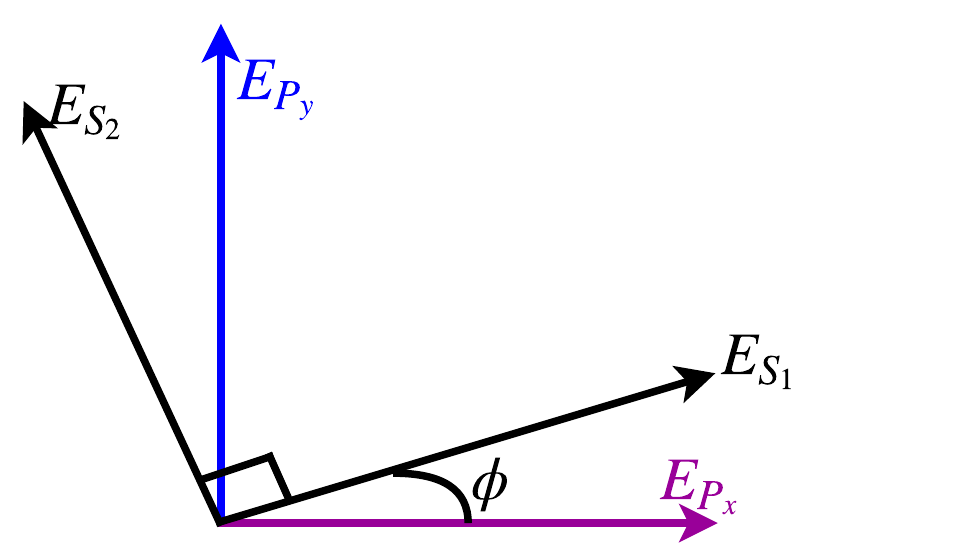}
		\label{vector}
	}
	\caption{\footnotesize(a) Schematic of the propagation direction and polarization of the pump and signal considering facet reflectivities. The dashed lines correspond to the reflected signal and pumps, (b) Vector diagram of the electric fields of counter-propagating orthogonally polarized degenerate pumps and polarization multiplexed signal.}
	\label{fig1}
\end{figure}

We consider polarization insensitive conjugate generation in an SOA through Bragg-scattering process. Polarization multiplexed signal is launched at the input port of the SOA and orthogonally polarized pumps are counter propagated (through input and output ports) in this configuration. Fig.~\ref{scheme1} shows the schematic of the propagation directions and the corresponding polarization of the pump and the signal in a scenario where the facet reflectivities of the SOA are also considered. 
Let the amplitudes of the orthogonally polarized pumps at the input of the SOA along the reference $X$ and $Y$ directions be $A_{P_X}$ and  $A_{P_Y}$ and that of the input signal  be represented as $\left(\left[A_{s1}\cos(\phi)-A_{s2}\sin(\phi)\right]\hat{a}_X+\left[A_{s1}\sin(\phi)+A_{s2}\cos(\phi)\right]\hat{a}_Y\right)$, where $A_{s1}$ and $A_{s2}$ denotes the orthogonal components of the complex amplitudes of the signal electric field, which is rotated with respect to the reference $X$ and $Y$ directions through an angle $\phi$ as shown in Fig.\ref{vector}.

\FloatBarrier

Thus the electric fields corresponding to the orthogonally polarized degenerate pumps at the respective inputs of the SOA can be represented as,
\begin{eqnarray}
\vec{E}_{P_x}=A_{P_x}\exp\left(j\left(\omega_pt-\phi_p\right)\right)\hat{a}_x,
\end{eqnarray}

\begin{eqnarray}
\vec{E}_{P_y}=A_{P_y}\exp\left(j\left(\omega_pt-\phi_p\right)\right)\hat{a}_y,
\end{eqnarray}
and that of the polarization multiplexed signal at the input is given by,
\begin{eqnarray}
\vec{E}_{S}=\exp\left(j\left(\omega_st-\phi_s\right)\right)\{\left[A_{s1}\cos(\phi)-A_{s2}\sin(\phi)\right]\hat{a}_x+\left[A_{s1}\sin(\phi)+A_{s2}\cos(\phi)\right]\hat{a}_y\},
\end{eqnarray}

where $\omega_p$ and $\omega_s$ represent the pump and signal frequencies respectively.\\


We first consider the case of negative detuning ($\omega_s<\omega_p$). Considering non-forbidden phase conjugate generation processes in the Bragg-scattering FWM in~\cite{Sobhanan}, the transfer matrix for the conjugate at both ports is given by, 

\begin{eqnarray}
\begin{bmatrix}
E_{cx}\\
E_{cy}\\
\end{bmatrix}=A^2_P  \exp\left(j(2\omega_p-\omega_s)t-(2\phi_p-\phi_s)\right)\eta\begin{bmatrix}
(\cos\phi & \pm\sin\phi\\
\cos\phi & \pm\sin\phi\\
\end{bmatrix}\begin{bmatrix}
A^*_{s1}\\
A^*_{s2}\\
\end{bmatrix}
\label{Eqn_new1}
\end{eqnarray}

Detailed derivation of Eqn.\ref{Eqn_new1} is given in Appendix.~\ref{app1}\newline

Thus for $\omega_s<\omega_p$, the transfer matrix corresponding to the conjugate generation process is non-unitary.\newline


We now consider the case of positive detuning. The transfer matrix of the conjugate, considering all the non-forbidden processes for output at both ports can be derived as,


\begin{eqnarray}
\begin{bmatrix}
E_{cx}\\
E_{cy}\\
\end{bmatrix}=A^2_P  \exp\left(j(2\omega_p-\omega_s)t-(2\phi_p-\phi_s)\right)\eta\begin{bmatrix}
\sin\phi & \cos\phi\\
\cos\phi & -\sin\phi\\
\end{bmatrix}\begin{bmatrix}
A^*_{s1}\\
A^*_{s2}\\
\end{bmatrix}
\label{Eqn_new2}
\end{eqnarray}
Detailed derivation of Eqn.\ref{Eqn_new2} is given in Appendix.~\ref{app2}\newline

Thus, the transfer matrix in case of $\omega_s>\omega_p$ is of the form of a unitary transformation.  $\eta$ represents the conjugate conversion efficiencies in each case, which is in-turn decided by the strength of the Bragg grating that contributes to the specific conjugate generation. We derive with appropriate approximations that polarization transformation for the positive detuning case is unitary while that of the negative detuning case is non-unitary. Unitary transformation is a desirable condition for the polarization demultiplexing algorithm in DSP. Hence the positive detuning is desirable for Bragg scattering-FWM aided polarization insensitive phase conjugate generation, both at port 1 and port 2 of the SOA.

\section{Conclusion}
The transfer function of polarization insensitive conjugate generation using counter-propagating orthogonally polarized pumps using Bragg scattering FWM in SOA in port 1 and port 2 is studied. The comparison is done between two cases $\omega_p>\omega_s$ and $\omega_s>\omega_p$. It is evident from this analysis that the optical phase conjugate process through Bragg-scattering FWM corresponds to a unitary transformation only for $\omega_s>\omega_p$ and the corresponding transformation is non-unitary for $\omega_s<\omega_p$. The efficiency of the process with $\omega_s>\omega_p$ is found to be smaller at least by 3 dB compared to that for $\omega_s<\omega_p$~\cite{Sobhanan}. In spite of poorer efficiency, a positive detuning is preferred because of the unitary nature of the corresponding transformation, and  consequently avoiding the limitations posed by digital signal processing in handling non-unitary transformation. This work leads to an insight of the selection of pump and signals frequency for the polarization insensitive conjugate generation process through Bragg-scattering FWM in SOA. 
\appendix
\section{Appendix A- $\omega_s<\omega_p$}\label{app1}
We consider all the non-forbidden processes that lead to phase conjugate generation. These processes are detailed in Tables. 3 \& 4 of ~\cite{Sobhanan} (reproduced here as Tables. 1 \& 2 for clarity). Forbidden processes are represented in red font and allowed processes are represented in green font.
\begin{table}[h!]
	\caption{\footnotesize{Frequencies generated in each port for cross-polarized and counter propagating pumps in SOA with non-zero facet reflectivity; Negative detuning  ($\omega_p>\omega_s$); grating due to co-propagating signal and pump. The signal is assumed to have arbitrary polarization. Frequencies shown in green satisfy phase matching condition}}
	\label{tab:negativeco}
	\begin{tabular}{ccccc}
		\toprule
		Interacting frequency & &\multicolumn{2}{c}{$\omega_p>\omega_s$ (Negative Detuning)}\\
		
		& &\multicolumn{2}{c}{Grating Frequency : $(\Omega=\omega_p-\omega_s)$}\\
		\midrule
		&\multicolumn{2}{c}{\textbf{Grating due to }\textbf{co-prop $\omega_s$ \& $\omega_p$}}&\multicolumn{2}{c}{\textbf{Grating due to} {\textbf{reflected $\omega_s$ \& $\omega_p$}}} \\
		&\multicolumn{2}{c}{\textbf{Forward}: Grating vector: $k_p-k_s$} & \multicolumn{2}{c}{\textbf{Backward}: Grating vector: $k_p-k_s$}\\
		\cmidrule(r){2-3}\cmidrule(l){4-5}
		& Port 1 & Port 2 & Port 1 & Port 2     \\
		\midrule
		$\omega_p$ (co-prop)-$x$  & $\omega_s$ & \textcolor{green}{$2\omega_p-\omega_s$ ($x$)} & \textcolor{red}{$2\omega_p-\omega_s$} & $\omega_s$  \\
		$\omega_p$ (reflected)-$y$  & $\omega_s$ & \textcolor{green}{$2\omega_p-\omega_s$ ($y$)} & \textcolor{red}{$2\omega_p-\omega_s$} & $\omega_s$  \\
		$\omega_p$ (counter-prop)- $y$ & $\omega_s$ & \textcolor{red}{$2\omega_p-\omega_s$} &\textcolor{green}{$2\omega_p-\omega_s$ ($y$)}  & $\omega_s$ \\
		$\omega_p$ (reflected)- $x$  & $\omega_s$ & \textcolor{red}{$2\omega_p-\omega_s$}& \textcolor{green}{$2\omega_p-\omega_s$ ($x$)} & $\omega_s$ \\
		$\omega_s$ (co-prop)- $x+y$  &\textcolor{red}{$2\omega_s-\omega_p$}  & $\omega_p$& $\omega_p$ & \textcolor{green}{ $2\omega_s-\omega_p$}\\
		$\omega_s$ (reflected)- $x+y$   & \textcolor{green}{$2\omega_s-\omega_p$} & $\omega_p$& $\omega_p$ &  \textcolor{red}{$2\omega_s-\omega_p$}\\
		\bottomrule
	\end{tabular}\\		
\end{table}
\begin{table}[h!]
	\caption{\footnotesize{Frequencies generated in each port for cross-polarized and counter propagating pumps in SOA with non-zero facet reflectivity; Negative detuning  ($\omega_p>\omega_s$); grating due to counter-propagating signal and pump. The signal is assumed to have arbitrary polarization. Frequencies shown in green satisfy phase matching condition}}
	\label{tab:negative_counter}
	\begin{tabular}{ccccc}
		\toprule
		Interacting frequency & &\multicolumn{2}{c}{$\omega_p>\omega_s$ (Negative Detuning)}\\
		
		& &\multicolumn{2}{c}{Grating Frequency : $(\Omega=\omega_p-\omega_s)$}\\
		\midrule
		&\multicolumn{2}{c}{\textbf{Grating due to }\textbf{reflected $\omega_s$ \&}}&\multicolumn{2}{c}{\textbf{Grating due to} {\textbf{co-prop $\omega_s$ \&}}} \\
		&\multicolumn{2}{c}{\textbf{co-propagating $\omega_p$}} &\multicolumn{2}{c}{\textbf{counter propagating $\omega_p$}} \\
		&\multicolumn{2}{c}{\textbf{Forward}: Grating vector: $k_p+k_s$} & \multicolumn{2}{c}{\textbf{Backward}: Grating vector: $k_p+k_s$}\\
		\cmidrule(r){2-3}\cmidrule(l){4-5}
		& Port 1 & Port 2 & Port 1 & Port 2     \\
		\midrule
		$\omega_p$ (co-prop)-$x$  & $\omega_s$ &\textcolor{red}{$2\omega_p-\omega_s$}  &\textcolor{green}{$2\omega_p-\omega_s$ ($x$)}  & $\omega_s$  \\
		$\omega_p$ (reflected)-$y$  & $\omega_s$ &\textcolor{red}{$2\omega_p-\omega_s$}  &\textcolor{green}{$2\omega_p-\omega_s$ ($y$)}  & $\omega_s$  \\
		$\omega_p$ (counter-prop)- $y$ & $\omega_s$ &\textcolor{green}{$2\omega_p-\omega_s$ ($y$)}   &\textcolor{red}{$2\omega_p-\omega_s$} & $\omega_s$ \\
		$\omega_p$ (reflected)- $x$  & $\omega_s$ &\textcolor{green}{$2\omega_p-\omega_s$ ($x$)} &\textcolor{red}{$2\omega_p-\omega_s$}  & $\omega_s$ \\
		$\omega_s$ (co-prop)- $x+y$  &\textcolor{green}{$2\omega_s-\omega_p$}  & $\omega_p$& $\omega_p$ & \textcolor{red}{ $2\omega_s-\omega_p$}\\
		$\omega_s$ (reflected)- $x+y$   & \textcolor{red}{$2\omega_s-\omega_p$} & $\omega_p$& $\omega_p$ &  \textcolor{green}{$2\omega_s-\omega_p$}\\
		\bottomrule
	\end{tabular}\\		
\end{table}

\subsection{Conjugate generation at port 1 of the SOA}
The primary contributions for the generation of conjugates in port 1 are due to the scattering of the (a) reflected pump and counter propagating pump off the  grating formed due to the beat between reflected signal and the counter propagating pump (with grating vector~$k_p-k_s$) and (b) co-propagating pump and reflected pump off the grating formed due to the beat between the co-propagating signal and counter propagating pump (with grating vector $k_p+k_s$). The facet reflectivities of the SOAs are typically small and hence the conjugate generated by scattering of reflected pumps can be neglected when compared to the efficiency of the conjugate generated by the transmitted pump.  The electric fields of conjugate generated through the remaining processes are written below, 

\begin{enumerate}
	\item  Conjugate generated in \textcolor{blue}{y-polarization} due to the interaction between counter propagating pump (y-polarization) with the backward propagating grating (grating vector :$k_p-k_s$, with a \textcolor{blue}{phase mismatch $\triangle k_c=0$}), represented as,
	\begin{eqnarray}
	\eta_1(\vec{E}_{P_y}.\vec{E}^*_{S})\vec{E}_{P_y}&=&\eta_1\left[\left(A_{P_y}\exp\left(j\left(\omega_pt-\phi_p\right)\right)\hat{a}_y\right).\left(\left[A^*_{s1}\cos\phi-A^*_{s2}\sin\phi\right]\hat{a}_x+\left[A^*_{s1}\sin\phi+A^*_{s2}\cos\phi\right]\hat{a}_y\right)\right] \nonumber\\
	& &A_{P_y}\exp\left(j\left(\omega_pt-\phi_p\right)\right)\hat{a}_y
	\nonumber \\
	&=&\eta_1 A_{P_y}^2 \exp\left(j(2\omega_p-\omega_s)t-(2\phi_p-\phi_s)\right) \left[A^*_{s1}\sin\phi+A^*_{s2}\cos\phi\right]\hat{a}_y
	\end{eqnarray}

	\item  Conjugate generated in \textcolor{magenta}{x-polarization} due to the interaction between co-propagating pump (x-polarization) with the backward propagating grating (grating vector :$k_p+k_s$, with a \textcolor{magenta}{phase mismatch $\triangle k_c=2(k_p-k_s)$}), represented as,
	\begin{eqnarray}
	\eta_2(\vec{E}_{P_y}.\vec{E}^*_{S})\vec{E}_{P_x}&=&\eta_2\left[\left(A_{P_y}\exp\left(j\left(\omega_pt-\phi_p\right)\right)\hat{a}_y\right).\left(\left[A^*_{s1}\cos\phi-A^*_{s2}\sin\phi\right]\hat{a}_x+\left[A^*_{s1}\sin\phi+A^*_{s2}\cos\phi\right]\hat{a}_y\right)\right] \nonumber \\
	& &A_{P_x}\exp\left(j\left(\omega_pt-\phi_p\right)\right)\hat{a}_x
	\nonumber \\
	&=&\eta_2 A_{P_y} A_{P_x} \exp\left(j(2\omega_p-\omega_s)t-(2\phi_p-\phi_s)\right) \left[A^*_{s1}\sin\phi+A^*_{s2}\cos\phi\right]\hat{a}_x
	\end{eqnarray}
	
\end{enumerate}
Here $\eta_1$ and $\eta_2$ represent the efficiency of each process.

The electric field of the Conjugate at Port 1, when orthogonal pumps carry equal power ($A_{P_x}=A_{P_y}=A_{P}$)

\begin{eqnarray}
E_C&=& \eta_1(\vec{E}_{P_y}.\vec{E}^*_{S})\vec{E}_{P_y}+  	\eta_3(\vec{E}_{P_y}.\vec{E}^*_{S})\vec{E}_{P_x}\nonumber \\
&=& A^2_P \exp\left(j(2\omega_p-\omega_s)t-(2\phi_p-\phi_s)\right)\left[A^*_{s1}\sin\phi+A^*_{s2}\cos\phi\right]
\left[\eta_2\hat{a}_x+\eta_1\hat{a}_y\right]
\end{eqnarray}
\begin{eqnarray}
\begin{bmatrix}
E_{cx}\\
E_{cy}\\
\end{bmatrix}=A^2_P  \exp\left(j(2\omega_p-\omega_s)t-(2\phi_p-\phi_s)\right)\begin{bmatrix}
\eta_2\sin\phi & \eta_2\cos\phi\\
\eta_1\sin\phi & \eta_1\cos\phi\\
\end{bmatrix}\begin{bmatrix}
A^*_{s1}\\
A^*_{s2}\\
\end{bmatrix}
\label{Eqn36}
\end{eqnarray}
\begin{eqnarray}
\begin{bmatrix}
E_{cx}\\
E_{cy}\\
\end{bmatrix}=A^2_P  \exp\left(j(2\omega_p-\omega_s)t-(2\phi_p-\phi_s)\right)\eta\begin{bmatrix}
\sin\phi & \cos\phi\\
\sin\phi & \cos\phi\\
\end{bmatrix}\begin{bmatrix}
A^*_{s1}\\
A^*_{s2}\\
\end{bmatrix},
\label{Eqn37}
\end{eqnarray}
where $\eta_1=\eta_2=\eta$, which can be achieved by adjusting the power levels  of the counter-propagating pumps.

\subsection{Conjugate generation at port 2 of the SOA}
The primary contributions for the generation of conjugates in port 2 are due to the scattering of the (a) reflected pump and co-propagating pump off the  grating formed due to the beat between co-propagating signal and pump (with grating vector $k_p-k_s$) and (b) counter-propagating pump and reflected pump off the grating formed due to the beat between the reflected signal and co-propagating pump (with grating vector $k_p+k_s$). The facet reflectivities of the SOAs are typically small and hence the conjugate generated by scattering of reflected pumps can be neglected while compared to the efficiency of the conjugate by the transmitted pump.  The electric fields of conjugate generated through the remaining processes are written below, 

\begin{enumerate}
	\item Conjugate generated in \textcolor{magenta} {x-polarization}, due to the interaction between co-propagating pump (x-polarization) with the forward propagating grating (grating vector: $k_p-k_s$, with a \textcolor{magenta}{phase mismatch $\triangle k_c=0$}), represented as,
	
	\begin{eqnarray}
	\eta_3(\vec{E}_{P_x}.\vec{E}^*_{S})\vec{E}_{P_x}&=&\eta_3\left[\left(A_{P_x}\exp\left(j\left(\omega_pt-\phi_p\right)\right)\hat{a}_x\right).\left(\left[A^*_{s1}\cos\phi-A^*_{s2}\sin\phi\right]\hat{a}_x+\left[A^*_{s1}\sin\phi+A^*_{s2}\cos\phi\right]\hat{a}_y\right)\right] \nonumber\\
	& &A_{P_x}\exp\left(j\left(\omega_pt-\phi_p\right)\right)\hat{a}_x
	\nonumber \\
	&=&\eta_3 A_{P_x}^2 \exp\left(j(2\omega_p-\omega_s)t-(2\phi_p-\phi_s)\right) \left[A^*_{s1}\cos\phi-A^*_{s2}\sin\phi\right]\hat{a}_x
	\end{eqnarray}
	
	\item Conjugate generated in \textcolor{blue}{y-polarization}, due to the interaction between counter-propagating pump (y-polarization) with the forward propagating grating (grating vector: $k_p+k_s$), with a \textcolor{blue}{phase mismatch $\triangle k_c=2(k_p-k_s)$}), represented as,

	\begin{eqnarray}
	\eta_4(\vec{E}_{P_x}.\vec{E}^*_{S})\vec{E}_{P_y}&=&\eta_4\left[\left(A_{P_x}\exp\left(j\left(\omega_pt-\phi_p\right)\right)\hat{a}_x\right).\left(\left[A^*_{s1}\cos\phi-A^*_{s2}\sin\phi\right]\hat{a}_x+\left[A^*_{s1}\sin\phi+A^*_{s2}\cos\phi\right]\hat{a}_y\right)\right] \nonumber\\
	& &A_{P_y}\exp\left(j\left(\omega_pt-\phi_p\right)\right)\hat{a}_y
	\nonumber \\
	&=&\eta_4 A_{P_x}A_{P_y} \exp\left(j(2\omega_p-\omega_s)t-(2\phi_p-\phi_s)\right) \left[A^*_{s1}\cos\phi-A^*_{s2}\sin\phi\right]\hat{a}_y
	\end{eqnarray}

\end{enumerate}
Here $\eta_3$ and $\eta_4$ represent the efficiency of each process.

The electric field of the conjugate at Port 2, when orthogonal pumps carry equal power ($A_{P_x}=A_{P_y}=A_{P}$), can be written as,

\begin{eqnarray}
E_C&=& \eta_3(\vec{E}_{P_y}.\vec{E}^*_{S})\vec{E}_{P_x}+ 	 \eta_4(\vec{E}_{P_x}.\vec{E}^*_{S})\vec{E}_{P_y}\nonumber \\
&=& A^2_P \exp\left(j(2\omega_p-\omega_s)t-(2\phi_p-\phi_s)\right)\left[A^*_{s1}\cos\phi-A^*_{s2}\sin\phi\right]
\left[\eta_3\hat{a}_x+(\eta_4\hat{a}_y\right]
\end{eqnarray}
\begin{eqnarray}
\begin{bmatrix}
E_{cx}\\
E_{cy}\\
\end{bmatrix}=A^2_P  \exp\left(j(2\omega_p-\omega_s)t-(2\phi_p-\phi_s)\right)\begin{bmatrix}
\eta_3\cos\phi & -\eta_3\sin\phi\\
\eta_4\cos\phi & -\eta_4\sin\phi\\
\end{bmatrix}\begin{bmatrix}
A^*_{s1}\\
A^*_{s2}\\
\end{bmatrix}
\label{Eqn42}
\end{eqnarray}

\begin{eqnarray}
\begin{bmatrix}
E_{cx}\\
E_{cy}\\
\end{bmatrix}=A^2_P  \exp\left(j(2\omega_p-\omega_s)t-(2\phi_p-\phi_s)\right)\eta\begin{bmatrix}
\cos\phi & -\sin\phi\\
\cos\phi & -\sin\phi\\
\end{bmatrix}\begin{bmatrix}
A^*_{s1}\\
A^*_{s2}\\
\end{bmatrix},
\label{Eqn43}
\end{eqnarray}
where $\eta_3=\eta_4=\eta$, which can be achieved by adjusting the power levels of the counter-propagating pumps.

\section{Appendix B- $\omega_s>\omega_p$}\label{app2}
In this section, we consider positive detuning case. We consider all the non-forbidden processes that lead to phase conjugate generation. These processes are detailed in Tables. 1 \& 2 of ~\cite{Sobhanan} (reproduced here as Tables. 3 \& 4 for clarity). Forbidden processes are represented in red font and allowed processes are represented in green font.
\begin{table}[h!]
	\caption{\footnotesize{Frequencies generated in each port of SOA with non-zero facet reflectivity; Positive detuning ($\omega_s>\omega_p$); grating due to co-propagating signal and pump. The signal is assumed to have arbitrary polarization. Pump and signal polarization and direction of propagation are indicated in the first column. Frequencies shown in green satisfy phase matching condition}}
	\label{tab:positive_co}
	\begin{tabular}{ccccc}
		\toprule
		Interacting frequency & &\multicolumn{2}{c}{$\omega_s>\omega_p$ (Positive Detuning)}\\
		
		& &\multicolumn{2}{c}{Grating Frequency : $(\Omega=\omega_s-\omega_p)$}\\
		\midrule
		&\multicolumn{2}{c}{\textbf{Grating due to }\textbf{co-prop $\omega_s$ \& $\omega_p$}}&\multicolumn{2}{c}{\textbf{Grating due to} {\textbf{reflected $\omega_s$ \& $\omega_p$}}} \\
		&\multicolumn{2}{c}{\textbf{Forward}; Grating vector: $k_s-k_p$} & \multicolumn{2}{c}{\textbf{Backward}; Grating vector: $k_s-k_p$}\\
		\cmidrule(r){2-3}\cmidrule(l){4-5}
		& Port 1 & Port 2 & Port 1 & Port 2     \\
		\midrule
		$\omega_p$ (co-prop)- $x$ & \textcolor{red}{$2\omega_p-\omega_s$} & $\omega_s$ & $\omega_s$ & \textcolor{green}{$2\omega_p-\omega_s$ ($x$)}  \\
		$\omega_p$ (reflected)- $y$ & \textcolor{red}{$2\omega_p-\omega_s$} & $\omega_s$ & $\omega_s$ & \textcolor{green}{$2\omega_p-\omega_s$ ($y$)}  \\
		$\omega_p$ (counter-prop)- $y$ & \textcolor{green}{$2\omega_p-\omega_s$ ($y$)} & $\omega_s$ & $\omega_s$ &\textcolor{red}{$2\omega_p-\omega_s$}  \\
		$\omega_p$ (reflected)- $x$ & \textcolor{green}{$2\omega_p-\omega_s$ ($x$)} & $\omega_s$& $\omega_s$ & \textcolor{red}{$2\omega_p-\omega_s$} \\
		$\omega_s$ (co-prop)- $\parallel+\perp$ & $\omega_p$ & \textcolor{green}{$2\omega_s-\omega_p$} & \textcolor{red} {$2\omega_s-\omega_p$} & $\omega_p$\\
		$\omega_s$ (reflected)- $\parallel+\perp$ & $\omega_p$ & \textcolor{red} {$2\omega_s-\omega_p$}& \textcolor{green}{$2\omega_s-\omega_p$} & $\omega_p$\\
		\bottomrule
	\end{tabular}\\		
\end{table}
\FloatBarrier
\begin{table}[h!]
	\caption{\footnotesize{Frequencies generated in each port of SOA with non-zero facet reflectivity; Positive detuning ($\omega_s>\omega_p$); grating due to counter-propagating signal and pump. The signal is assumed to have arbitrary polarization.Frequencies shown in green satisfy phase matching condition}}
	\label{tab:positive_counter}
	\begin{tabular}{ccccc}
		\toprule
		Interacting frequency & &\multicolumn{2}{c}{$\omega_s>\omega_p$ (Positive Detuning)}\\
		
		& &\multicolumn{2}{c}{Grating Frequency : $(\Omega=\omega_s-\omega_p)$}\\
		\midrule
		&\multicolumn{2}{c}{\textbf{Grating due to} \textbf{co-prop $\omega_s$ \&}}&\multicolumn{2}{c}{\textbf{Grating due to reflected $\omega_s$ \&}}\\
		&\multicolumn{2}{c}{\textbf{counter propagating $\omega_p$}} &\multicolumn{2}{c}{\textbf{co-propagating $\omega_p$}} \\
		&\multicolumn{2}{c}{\textbf{Forward}; Grating vector: $k_s+k_p$} & \multicolumn{2}{c}{\textbf{Backward}; Grating vector: $k_s+k_p$}\\
		\cmidrule(r){2-3}\cmidrule(l){4-5}
		& Port 1 & Port 2 & Port 1 & Port 2     \\
		\midrule
		$\omega_p$ (co-prop)- $x$ & \textcolor{green}{$2\omega_p-\omega_s$ ($x$)} & $\omega_s$ & $\omega_s$ & \textcolor{red}{$2\omega_p-\omega_s$}  \\
		$\omega_p$ (reflected)- $y$ & \textcolor{green}{$2\omega_p-\omega_s$ ($y$)} & $\omega_s$ & $\omega_s$ & \textcolor{red}{$2\omega_p-\omega_s$}  \\
		$\omega_p$ (counter-prop)- $y$ & \textcolor{red}{$2\omega_p-\omega_s$} & $\omega_s$ & $\omega_s$ &\textcolor{green}{$2\omega_p-\omega_s$ ($y$)}  \\
		$\omega_p$ (reflected)- $x$ & \textcolor{red}{$2\omega_p-\omega_s$} & $\omega_s$& $\omega_s$ & \textcolor{green}{$2\omega_p-\omega_s$ ($x$)} \\
		$\omega_s$ (co-prop)- $\parallel+\perp$ & $\omega_p$ & \textcolor{red}{$2\omega_s-\omega_p$} & \textcolor{green} {$2\omega_s-\omega_p$} & $\omega_p$\\
		$\omega_s$ (reflected)- $\parallel+\perp$ & $\omega_p$ & \textcolor{green} {$2\omega_s-\omega_p$}& \textcolor{red}{$2\omega_s-\omega_p$} & $\omega_p$\\
		\bottomrule
	\end{tabular}\\		
	\end{table}
\subsection{Conjugate generation at port 1 of the SOA}
The primary contributions for the generation of conjugates in port 1 are due to the scattering of the (a) reflected pump and counter propagating pump off the  grating formed due to the beat between co-propagating signal and pump (with grating vector~$k_s-k_p$) and (b) co-propagating pump and reflected pump off the grating formed due to the beat between the co-propagating signal and counter propagating pump (with grating vector $k_s+k_p$). Considering low facet reflectivities of SOA, we assume that, the conjugate power due to reflected pumps are comparatively smaller and are hence neglected. The electric fields of conjugate generated through the remaining processes are written as,

\begin{enumerate}
	\item Conjugate generated in (\textcolor{blue}{y-polarization}) due to the interaction between counter-propagating pump (y-polarization) with the forward propagating grating (grating vector: $k_s-k_p$, with a \textcolor{blue}{phase mismatch $\triangle k_c=2(k_p-k_s)$}), represented as,
	
	\begin{eqnarray}
	\eta_1(\vec{E}_{P_x}.\vec{E}^*_{S})\vec{E}_{P_y}&=&\eta_1\left[\left(A_{P_x}\exp\left(j\left(\omega_pt-\phi_p\right)\right)\hat{a}_x\right).\left(\left[A^*_{s1}\cos\phi-A^*_{s2}\sin\phi\right]\hat{a}_x+\left[A^*_{s1}\sin\phi+A^*_{s2}\cos\phi\right]\hat{a}_y\right)\right] \nonumber\\
	& &A_{P_y}\exp\left(j\left(\omega_pt-\phi_p\right)\right)\hat{a}_y
	\nonumber \\
	&=&\eta_1 A_{P_x}A_{P_y} \exp\left(j(2\omega_p-\omega_s)t-(2\phi_p-\phi_s)\right) \left[A^*_{s1}\cos\phi-A^*_{s2}\sin\phi\right]\hat{a}_y
	\end{eqnarray}

	\item  Conjugate generated in (\textcolor{magenta}{x-polarization}) due to the interaction between co-propagating pump (x-polarization) with the forward propagating grating (grating vector: $k_s+k_p$, with a \textcolor{magenta}{phase mismatch $\triangle k_c=2(k_p-k_s)$}), represented as,
	
	\begin{eqnarray}
	\eta_2(\vec{E}_{P_y}.\vec{E}^*_{S})\vec{E}_{P_x}&=&\eta_2\left[\left(A_{P_y}\exp\left(j\left(\omega_pt-\phi_p\right)\right)\hat{a}_y\right).\left(\left[A^*_{s1}\cos\phi-A^*_{s2}\sin\phi\right]\hat{a}_x+\left[A^*_{s1}\sin\phi+A^*_{s2}\cos\phi\right]\hat{a}_y\right)\right] \nonumber \\
	& &A_{P_x}\exp\left(j\left(\omega_pt-\phi_p\right)\right)\hat{a}_x
	\nonumber \\
	&=&\eta_2 A_{P_y} A_{P_x} \exp\left(j(2\omega_p-\omega_s)t-(2\phi_p-\phi_s)\right) \left[A^*_{s1}\sin\phi+A^*_{s2}\cos\phi\right]\hat{a}_x
	\end{eqnarray}

\end{enumerate}
The electric field of the conjugate at Port 1, when orthogonal pumps carry equal power ($A_{P_x}=A_{P_y}=A_{P}$)

\begin{eqnarray}
E_C&=& \eta_1(\vec{E}_{P_x}.\vec{E}^*_{S})\vec{E}_{P_y}+  \eta_2(\vec{E}_{P_y}.\vec{E}^*_{S})\vec{E}_{P_x}\nonumber \\
&=& A^2_P  \exp\left(j(2\omega_p-\omega_s)t-(2\phi_p-\phi_s)\right)\nonumber\\
& &[\left(\eta_2\left[A^*_{s1}\sin\phi+A^*_{s2}\cos\phi\right]\hat{a}_x\right)+\left(\eta_1\left[A^*_{s1}\cos\phi-A^*_{s2}\sin\phi\right]\right)\hat{a}_y]
\end{eqnarray}

\begin{eqnarray}
\begin{bmatrix}
E_{cx}\\
E_{cy}\\
\end{bmatrix}=A^2_P  \exp\left(j(2\omega_p-\omega_s)t-(2\phi_p-\phi_s)\right)\begin{bmatrix}
\eta_2\sin\phi & \eta_2\cos\phi\\
\eta_1\cos\phi & -\eta_1\sin\phi\\
\end{bmatrix}\begin{bmatrix}
A^*_{s1}\\
A^*_{s2}\\
\end{bmatrix}
\label{Eqn48}
\end{eqnarray}\\

\begin{eqnarray}
\begin{bmatrix}
E_{cx}\\
E_{cy}\\
\end{bmatrix}=A^2_P  \exp\left(j(2\omega_p-\omega_s)t-(2\phi_p-\phi_s)\right)\eta\begin{bmatrix}
\sin\phi & \cos\phi\\
\cos\phi & -\sin\phi\\
\end{bmatrix}\begin{bmatrix}
A^*_{s1}\\
A^*_{s2}\\
\end{bmatrix},
\label{Eqn49}
\end{eqnarray}\\
where, $\eta_1=\eta_2=\eta$, which can be achieved by adjusting the power levels of the counter-propagating pumps.
\subsection{Conjugate generation at port 2 of the SOA}
The primary contributions for the generation of conjugates in port 2 are due to the scattering of the (a) reflected pump and co-propagating pump off the  grating formed due to the beat between reflected signal and counter propagating pump (with grating vector $k_s-k_p$) and (b) counter-propagating pump and reflected pump off the grating formed due to the beat between the reflected signal and co-propagating pump (with grating vector $k_s+k_p$). Here again, we assume that the conjugation due to  reflected pump is negligible. The electric fields of conjugate generated through the remaining processes can be written as,

\begin{enumerate}
	\item  Conjugate generated in (\textcolor{magenta}{x-polarization}) due to the interaction between co-propagating pump (x-polarization) with the backward propagating grating (grating vector: $k_s-k_p$, with a \textcolor{magenta}{phase mismatch $\triangle k_c=2(k_p-k_s)$}), represented as,
	
	\begin{eqnarray}
	\eta_3(\vec{E}_{P_y}.\vec{E}^*_{S})\vec{E}_{P_x}&=&\eta_3\left[\left(A_{P_y}\exp\left(j\left(\omega_pt-\phi_p\right)\right)\hat{a}_y\right).\left(\left[A^*_{s1}\cos\phi-A^*_{s2}\sin\phi\right]\hat{a}_x+\left[A^*_{s1}\sin\phi+A^*_{s2}\cos\phi\right]\hat{a}_y\right)\right] \nonumber \\
	& &A_{P_x}\exp\left(j\left(\omega_pt-\phi_p\right)\right)\hat{a}_x
	\nonumber \\
	&=&\eta_3 A_{P_y} A_{P_x} \exp\left(j(2\omega_p-\omega_s)t-(2\phi_p-\phi_s)\right) \left[A^*_{s1}\sin\phi+A^*_{s2}\cos\phi\right]\hat{a}_x
	\end{eqnarray}

	\item  Conjugate generated in (\textcolor{blue}{y-polarization}) due to the interaction between counter propagating pump (y-polarization) with the backward propagating grating (grating vector: $k_s+k_p$, with a \textcolor{blue}{phase mismatch $\triangle k_c=2(k_p-k_s)$}), represented as,
	
	\begin{eqnarray}
	\eta_4(\vec{E}_{P_x}.\vec{E}^*_{S})\vec{E}_{P_y}&=&\eta_4\left[\left(A_{P_x}\exp\left(j\left(\omega_pt-\phi_p\right)\right)\hat{a}_x\right).\left(\left[A^*_{s1}\cos\phi-A^*_{s2}\sin\phi\right]\hat{a}_x+\left[A^*_{s1}\sin\phi+A^*_{s2}\cos\phi\right]\hat{a}_y\right)\right] \nonumber\\
	& &A_{P_y}\exp\left(j\left(\omega_pt-\phi_p\right)\right)\hat{a}_y
	\nonumber \\
	&=&\eta_4 A_{P_x}A_{P_y} \exp\left(j(2\omega_p-\omega_s)t-(2\phi_p-\phi_s)\right) \left[A^*_{s1}\cos\phi-A^*_{s2}\sin\phi\right]\hat{a}_y
	\end{eqnarray}

\end{enumerate}
Conjugate at Port 2, when orthogonal pumps carry equal power ($A_{P_x}=A_{P_y}=A_{P}$)
\begin{eqnarray}
E_C&=& \eta_3(\vec{E}_{P_y}.\vec{E}^*_{S})\vec{E}_{P_x}+ 	 \eta_4(\vec{E}_{P_x}.\vec{E}^*_{S})\vec{E}_{P_y}\nonumber \\
&=& A^2_P  \exp\left(j(2\omega_p-\omega_s)t-(2\phi_p-\phi_s)\right)\nonumber\\
& &[\left(\eta_3\left[A^*_{s1}\sin\phi+A^*_{s2}\cos\phi\right]\hat{a}_x\right)+\left(\eta_4\left[A^*_{s1}\cos\phi-A^*_{s2}\sin\phi\right]\right)\hat{a}_y]
\end{eqnarray}
\begin{eqnarray}
\begin{bmatrix}
E_{cx}\\
E_{cy}\\
\end{bmatrix}=A^2_P  \exp\left(j(2\omega_p-\omega_s)t-(2\phi_p-\phi_s)\right)\begin{bmatrix}
\eta_3\sin\phi & \eta_3\cos\phi\\
\eta_4\cos\phi & -\eta_4\sin\phi\\
\end{bmatrix}\begin{bmatrix}
A^*_{s1}\\
A^*_{s2}\\
\end{bmatrix}
\label{Eqn54}
\end{eqnarray}

\begin{eqnarray}
\begin{bmatrix}
E_{cx}\\
E_{cy}\\
\end{bmatrix}=A^2_P  \exp\left(j(2\omega_p-\omega_s)t-(2\phi_p-\phi_s)\right)\eta\begin{bmatrix}
\sin\phi & \cos\phi\\
\cos\phi & -\sin\phi\\
\end{bmatrix}\begin{bmatrix}
A^*_{s1}\\
A^*_{s2}\\
\end{bmatrix},
\label{Eqn55}
\end{eqnarray}
where, $\eta_3=\eta_4=\eta$, which can be achieved by adjusting the power levels of the counter-propagating pumps.
 

\end{document}